# Characterizing the NP-PSPACE Gap in the Satisfiability Problem for Modal Logic[*]

Joseph Y. Halpern
Computer Science Department
Cornell University, U.S.A.
e-mail: halpern@cs.cornell.edu

Leandro Chaves Rêgo
School of Electrical and Computer Engineering
Cornell University, U.S.A.
e-mail: lcr26@cornell.edu

February 5, 2018

### Abstract

There has been a great of work on characterizing the complexity of the satisfiability and validity problem for modal logics. In particular, Ladner showed that the validity problem for all logics between K, T, and S4 is *PSPACE*-complete, while for S5 it is *NP*-complete. We show that, in a precise sense, it is *negative introspection*, the axiom $\neg Kp \Rightarrow K\neg Kp$, that causes the gap. In a precise sense, if we require this axiom, then the satisfiability problem is *NP*-complete; without it, it is *PSPACE*-complete.

## 1 Introduction

There has been a great of work on characterizing the complexity of the satisfiability and validity problem for modal logics (see [Halpern and Moses 1992; Ladner 1977; Vardi 1989] for some examples of most interest here). In particular, Ladner [1977] showed that the validity problem for all logics between K, T, and S4 is *PSPACE*-complete, while for S5 it is *NP*-complete.

[*]This work was supported in part by NSF under grants CTC-0208535, ITR-0325453, and IIS-0534064, by ONR under grants N00014-00-1-03-41 and N00014-01-10-511, and by the DoD Multidisciplinary University Research Initiative (MURI) program administered by the ONR under grant N00014-01-1-0795. The second author was also supported in part by a scholarship from the Brazilian Government through the Conselho Nacional de Desenvolvimento Científico e Tecnológico (CNPq).

An easy modification of his argument shows that it is *NP*-complete for KD45 as well. There has been followup work trying to understand what causes the gap between *NP* and *PSPACE*. Expressed in terms of epistemic reasoning, Vardi [1989] showed that, in a precise sense, the ability to combine distinct items of knowledge, as characterized by the axiom $Kp \land Kq \Rightarrow K(p \land q)$, causes the gap; requiring this axiom gives *PSPACE*-completeness, while without it we get *NP*-completeness. We give another characterization. We show that, in a precise sense, it is *negative introspection*, the axiom $\neg Kp \Rightarrow K \neg Kp$, that causes the gap. In a precise sense, if we require this axiom, then the satisfiability problem is *NP*-complete; without it, it is *PSPACE*-complete.

More precisely, consider the following axioms and inference rules, all of which have been well-studied in the literature [Fagin, Halpern, Moses, and Vardi 1995]:

A1. All tautologies of propositional calculus

A2. $(K\varphi \land K(\varphi \Rightarrow \psi)) \Rightarrow K\psi$ (Distribution Axiom)

A3. $K\varphi \Rightarrow \varphi$ (Knowledge Axiom)

A4. $K\varphi \Rightarrow KK\varphi$ (Positive Introspection Axiom)

A5. $\neg K\varphi \Rightarrow K\neg K\varphi$, (Negative Introspection Axiom)

A6. $\neg K(false)$ (Consistency Axiom)

R1. From $\varphi$ and $\varphi \Rightarrow \psi$ infer $\psi$ (modus ponens)

R2. From $\varphi$ infer $K\varphi$, (Knowledge Generalization)

The system K consists of A1, A2, R1, and R2; the system T is K + A3; the system A4 is T + A4; the system S5 is S4 + A5; the system KD45 is S5 + A6 − A3.

Consider any modal logic that includes A1, A2, R1, R2 and some (possibly empty) subset of A3, A4, and A5. We show that the satisfiability for the resulting logic is *NP*-complete iff A5 is included; otherwise it is *PSPACE*-complete. While this result follows easily from well-known techniques, it does not seem to have been observed before.

## 2 Modal Logic: A Brief Review

We briefly review basic modal logic, introducing the notation used in the statement and proof of our result. The syntax of the modal logic is as follows: formulas are formed by starting with a set $\Phi = \{p, q, \ldots\}$ of primitive propositions, and then closing off under conjunction ($\land$), negation ($\neg$), and the modal operator $K$. Call the resulting language $\mathcal{L}_1^K(\Phi)$. As usual, we define $\varphi \lor \psi$ and $\varphi \Rightarrow \psi$ as abbreviations of $\neg(\neg\varphi \land \neg\psi)$ and $\neg\varphi \lor \psi$, respectively. The intended interpretation of $K\varphi$ varies depending on the context. It typically has been interpreted as knowledge, as belief, and as necessity. Under the epistemic interpretation, $K\varphi$ is read as "the agent *knows* $\varphi$"; under the necessity interpretation, $K\varphi$ can be read "$\varphi$ is necessarily true".

The standard approach to giving semantics to formulas in $\mathcal{L}_1^K(\Phi)$ is by means of Kripke structures. A tuple $M = (S, \pi, \mathcal{K})$ is a *Kripke structure (over $\Phi$)* if $S$ is a set of states, $\pi : S \times \Phi \to \{\mathbf{true}, \mathbf{false}\}$ is an *interpretation* that determines which primitive propositions are true at each state, $\mathcal{K}$ is a binary relation on $S$. Intuitively, $(s,t) \in \mathcal{K}$ if, in state $s$, state $t$ is considered possible (by the agent, if we are thinking of $K$ as representing an agent's knowledge or belief). For convenience, we define $\mathcal{K}(s) = \{t : (s,t) \in \mathcal{K}\}$.

Let $\mathcal{M}(\Phi)$ denote the class of all Kripke structures over $\Phi$ with no restrictions on the $\mathcal{K}$ relation. Depending on the desired interpretation of the formula $K\varphi$, a number of conditions may be imposed on the binary relation $\mathcal{K}$. $\mathcal{K}$ is *reflexive* if for all $s \in S$, $(s,s) \in \mathcal{K}$; it is *transitive* if for all $s,t,u \in S$, if $(s,t) \in \mathcal{K}$ and $(t,u) \in \mathcal{K}$, then $(s,u) \in \mathcal{K}$; it is *serial* if for all $s \in S$ there exists $t \in S$ such that $(s,t) \in \mathcal{K}$; it is *Euclidean* if for all $s,t,u \in S$, if $(s,t) \in \mathcal{K}$ and $(s,u) \in \mathcal{K}$ then $(t,u) \in \mathcal{K}$. We use the superscripts $r$, $e$, $t$ and $s$ to indicate that the $\mathcal{K}$ relation is restricted to being reflexive, Euclidean, transitive, and serial, respectively. Thus, for example, $\mathcal{M}^{rt}(\Phi)$ is the class of all Kripke strutures where the $\mathcal{K}$ relation is reflexive and transitive.

We write $(M,s) \models \varphi$ if $\varphi$ is true at state $s$ in the Kripke structure $M$. The truth relation is defined inductively as follows:

$$(M,s) \models p, \text{ for } p \in \Phi, \text{ if } \pi(s,p) = \mathbf{true}$$
$$(M,s) \models \neg\varphi \text{ if } (M,s) \not\models \varphi$$
$$(M,s) \models \varphi \wedge \psi \text{ if } (M,s) \models \varphi \text{ and } (M,s) \models \psi$$
$$(M,s) \models K\varphi \text{ if } (M,t) \models \varphi \text{ for all } t \text{ such that } (s,t) \in \mathcal{K}$$

A formula $\varphi$ is said to be *satisfiable in Kripke structure $M$* if there exists $s \in S$ such that $(M,s) \models \varphi$; $\varphi$ *valid in $M$* if $(M,s) \models \varphi$ for all $s \in S$. A formula is *satisfiable* (resp., *valid*) *in a class $\mathcal{N}$* of Kripke structures if it is satisfiable in some Kripke structure in $\mathcal{N}$ (resp., valid in all Kripke structures in $\mathcal{N}$).

There is a well-known correspondence between properties of the $\mathcal{K}$ relation and axioms: reflexivity corresponds to A3, transitivity corresponds to A4, the Euclidean property corresponds to A5, and the serial property corresponds to A6. This correspondence is made precise in the following theorem.

**Theorem 2.1:** [Fagin, Halpern, Moses, and Vardi 1995] *Let $\mathcal{C}$ be a (possibly empty) subset of $\{A3, A4, A5, A6\}$ and let $C$ be the corresponding subset of $\{r,t,e,s\}$. Then $\{A1, A2, R1, R2\} \cup \mathcal{C}$ is a sound and complete axiomatization of the language $\mathcal{L}_1^K(\Phi)$ with respect to $\mathcal{M}^C(\Phi)$.*

## 3 The Result

We can now state our main result.

**Theorem 3.1:** *For $C \subseteq \{r, e, t, s\}$, the complexity of the problem of deciding if a formula $\varphi \in \mathcal{L}_1^K(\Phi)$ is satisfiable in $\mathcal{M}^C(\Phi)$ is NP-complete if $e \in C$, and is PSPACE-complete if $e \notin C$.*

The theorem claimed in the introduction, namely, that the satisfiability problem for the logics discussed in the introduction is *NP*-complete iff A5 is an axiom, and otherwise is *PSPACE*-complete, follows immediately from Theorems 2.1 and 3.1.

**Proof:** Much of the proof of Theorem 3.1 is known. In particular, the *PSPACE* hardness result in the case that $e \notin C$ follows from Ladner's results. Ladner proves the matching upper bound if $C = \emptyset$ (the system K), $C = \{r\}$ (the system T), and $C = \{r, t\}$ (the system S4). Since a reflexive relation is serial, Ladner's results deal with the cases $C = \{r, s\}$ and $C = \{r, t, s\}$ as well. It is straightforward to modify Ladner's argument to get *PSPACE*-completeness in the remaining cases where $e \notin C$. For the cases where $e \in C$, it is well known (and easy to show) that if a relation is reflexive and Euclidean, then it is symmetric, transitive, and serial, so $\mathcal{M}^{r,e} = \mathcal{M}^{r,e,t} = \mathcal{M}^{r,e,s} = \mathcal{M}^{r,e,s,t}$. Since Ladner proves *NP*-completeness for S5, when the $\mathcal{K}$ relation is an equivalence relation, *NP*-completeness in all these cases follows. *NP*-completeness in the case where $C = \{r, t\}$ is proved in [Fagin, Halpern, Moses, and Vardi 1995], using a slight modification of Ladner's techniques. We generalize these arguments to deal with the case that $C = \{e\}$. All the remaining cases actually follow from our argument.

As usual, let $|\varphi|$ denote the length of $\varphi$ when viewed a string of symbols. As in the proof of Proposition 3.6.2 in [Fagin, Halpern, Moses, and Vardi 1995], the key step in showing *NP*-completness lies in showing that a formula is satisfiable in $\mathcal{M}^e$ iff it is satisfiable in a structure with few states of a particular type. This characterization (and its proof) is a generalization of analogous characterizations given in Propositions 3.1.6 and 3.6.2 in [Fagin, Halpern, Moses, and Vardi 1995] for satisfiability in $\mathcal{M}^{r,e,t}$ and $\mathcal{M}^{e,s,t}$.

**Lemma 3.2:** *A formula $\varphi$ is satisfiable in $\mathcal{M}^e$ iff there exists some structure $M$ such that $(M, s_0) \models \varphi$, where $M = (\{s_0\} \cup S \cup S', \pi, \mathcal{K})$, and (a) $S$ and $S'$ are disjoint sets of states; (b) if $S = \emptyset$ then $S' = \emptyset$, (c) $\mathcal{K}(s_0) = S$; (d) $\mathcal{K}(s) = S \cup S'$ if $s \in S \cup S'$; and (e) $|\{s_0\} \cup S \cup S'| \leq |\varphi|$.*

Note that we may have $s_0 \notin S \cup S'$. However, if $s_0 \in S \cup S'$, then it follows from conditions (a), (c), and (d) that $S' = \emptyset$. We get a characterization for

- $\mathcal{M}^{es}$ by requiring that $S \neq \emptyset$;
- $\mathcal{M}^{et}$ by requiring that $S' = \emptyset$;
- $\mathcal{M}^{est}$ by requiring that $S' = \emptyset$ and $S \neq \emptyset$;
- $\mathcal{M}^{re} = \mathcal{M}^{ret} = \mathcal{M}^{rest}$ by requiring that $s_0 \in S$ (so that, as we have observed, $S' = \emptyset$).

The third and fourth characterizations are Proposition 3.1.6 in [Fagin, Halpern, Moses, and Vardi 1995].

**Proof:** Suppose that $M$ is a structure of the type described in the statement of the lemma, $(s,t) \in \mathcal{K}$, and $(s,t') \in \mathcal{K}$. Then $t' \in \mathcal{K}(t) = S \cup S'$, so $\mathcal{K}$ is Euclidean. It follows that $M \in \mathcal{M}^e$. Thus, if $(M, s_0) \models \varphi$, then $\varphi$ is satisfiable in $\mathcal{M}^e$.

For the converse, we proceed much as in the proof of Propositions 3.1.6 and 3.6.2 in [Fagin, Halpern, Moses, and Vardi 1995]. Suppose that $M' = (T, \pi', \mathcal{K}') \in \mathcal{M}^e$, $s_0 \in T$, and $(M', s_0) \models \varphi$. Let $F_1$ be the set of subformulas of $\varphi$ of the form $K\psi$ such that $(M', s_0) \models \neg K\psi$, and let $F_2$ be the set of subformulas of $\varphi$ of the form $K\psi$ such that $KK\psi$ is a subformula of $\varphi$ and $(M', s_0) \models \neg KK\psi \wedge K\psi$. (We remark that it is not hard to show that if $M \in \mathcal{M}^C$ where $e$ and at least one of $r$ or $t$ is in $C$, then $F_2 = \emptyset$.)

For each formula $K\psi \in F_1$, there must exist a state $s_\psi^{F_1} \in \mathcal{K}'(s_0)$ such that $(M', s_\psi^{F_1}) \models \neg\psi$. Note that if $F_1 \neq \emptyset$ then $\mathcal{K}'(s_0) \neq \emptyset$. Define $I(s_0) = \{s_0\}$ if $s_0 \in \mathcal{K}'(s_0)$, and $I(s_0) = \emptyset$ otherwise. Let $S = \{s_\psi^{F_1} : K\psi \in F_1\} \cup I(s_0)$. If $K\psi \in F_2$ then $KK\psi \in F_1$, so there must exist a state $s_\psi^{F_2} \in \mathcal{K}'(s_{K\psi}^{F_1})$ such that $(M', s_\psi^{F_2}) \models \neg\psi$. Moreover, since $(M', s_0) \models K\psi$, it must be the case that $s_\psi^{F_2} \notin \mathcal{K}'(s_0)$. Let $S'' = \{s_\psi^{F_2} : K\psi \in F_2\}$. By construction, $S'$ and $S''$ are disjoint. Moreover, if $S = \emptyset$, then $F_1 = \emptyset$, so $F_2 = \emptyset$ and $S' = \emptyset$.

Let $S^* = \{s_0\} \cup S \cup S'$. Define the binary relation $\mathcal{K}$ on $S^*$ by taking $\mathcal{K}(s_0) = S$ and $\mathcal{K}(t) = S \cup S'$ for $t \in S \cup S'$. To show that this is well defined, we must show that (a) $s_0 \notin S'$ and that (b) if $s_0 \in S$, then $S' = \emptyset$. For (a), suppose by way of contradiction that $s_0 \in S'$. Thus, there exists $s \in S$ such that $s_0 \in \mathcal{K}'(s)$. By the Euclidean property, it follows that $s_0 \in \mathcal{K}'(s_0)$, a contradiction since $S'$ is disjoint from $\mathcal{K}'(s_0)$. For (b), note that if $s_0 \in S$, then $s_0 \in \mathcal{K}'(s_0)$. It is easy to see that if $s, s' \in \mathcal{K}'(s_0)$, then $\mathcal{K}'(s) = \mathcal{K}'(s')$. For if $s, s' \in \mathcal{K}'(s_0)$ then, by the Euclidean property, $s' \in \mathcal{K}'(s)$. Thus, if $t \in \mathcal{K}'(s)$, another application of the Euclidean property shows that $t \in \mathcal{K}'(s')$. Hence, $\mathcal{K}'(s') \subseteq \mathcal{K}'(s)$. A symmetric argument gives equality. But now suppose that $t \in S'$. Then, as we have observed, there exists some $s \in S$ such that $t \in \mathcal{K}'(s) - \mathcal{K}'(s_0)$. But if $s_0 \in S$, then $\mathcal{K}'(s) - \mathcal{K}'(s_0) = \emptyset$. Thus, $S' = \emptyset$ if $s_0 \in S$.

A similar argument shows that $\mathcal{K}$ is the restriction of $\mathcal{K}'$ to $S^*$. For clearly $S'$ is disjoint from $\mathcal{K}'(s_0)$, so $\mathcal{K}(s_0) = \mathcal{K}'(s_0) \cap S^*$. Now suppose that $s \in S \cup S'$. It is easy to see that there exists some $s' \in S$ such that $s \in \mathcal{K}'(s')$. This is clear by construction if $s \in S'$. And if $s \in S$, then $s \in \mathcal{K}'(s_0)$ and, by the Euclidean property, $s \in \mathcal{K}'(s)$. If $t \in S \cup S'$, we want to show that $t \in \mathcal{K}'(s)$. Again, there exists some $t'$ such that $t' \in S$ and $t \in \mathcal{K}'(t')$. Since $s', t' \in \mathcal{K}'(s_0)$, by the Euclidean property, $s' \in \mathcal{K}'(t')$. Since $s', t \in \mathcal{K}'(t')$, the Euclidean property implies that $t \in \mathcal{K}'(s')$. Since $s, t \in \mathcal{K}'(s')$, yet another application of the Euclidean property shows that $t \in \mathcal{K}'(s)$. Thus, $\mathcal{K}(s) \subseteq \mathcal{K}'(s) \cap S^*$. To prove equality suppose that $t \in \mathcal{K}'(s) \cap S^*$. If $t \in S \cup S'$, then by definition $t \in \mathcal{K}(s)$. If $t = s_0$, then by the Euclidean property it follows that $s_0 \in \mathcal{K}'(s_0)$, so $s_0 \in S \subseteq \mathcal{K}(s)$. Thus, $t \in \mathcal{K}(s)$, as desired.

Let $M = (S^*, \pi, \mathcal{K})$, where $\pi$ is the restriction of $\pi'$ to $\{s_0\} \cup S \cup S'$. It is well known [Fagin, Halpern, Moses, and Vardi 1995] that there are at most $|\varphi|$ subformulas of $\varphi$. Since $F_1$ and $F_2$ are disjoint sets of subformulas of $\varphi$, form $K\psi$, and at least one subformula of $\varphi$ is a primitive proposition (and thus not of the form $K\psi$), it must be the case that $|F_1| + |F_2| \leq |\varphi| - 1$, giving us the desired bound on the number of states.

We now show that for all states $s \in S^*$ and for all subformulas $\psi$ of $\varphi$ (including $\varphi$ itself), $(M, s) \models \psi$ iff $(M', s) \models \psi$. The proof proceeds by induction on the structure of $\varphi$. The only nontrivial case is when $\psi$ is of the form $K\psi'$. If $(M', s) \models K\psi'$, then $(M', t) \models \psi'$ for all $t \in \mathcal{K}'(t)$. Since $\mathcal{K}$ is the restriction of $\mathcal{K}'$ to $S^*$, this implies that $(M', t) \models \psi'$ for all $t \in \mathcal{K}(s)$. Thus, by the induction hypothesis, $(M, t) \models \psi'$ for all $t \in \mathcal{K}(s)$; that is, $(M, s) \models K\psi'$. For the converse, suppose that $(M', s) \models \neg K\psi'$. If it is also the case that $(M', s_0) \models \neg K\psi'$, then $K\psi' \in F_1$. By the induction hypothesis, $(M, s_{\psi'}^{F_1}) \models \neg \psi'$. Thus, $(M, s) \models \neg K\psi'$. If $(M', s_0) \models K\psi'$, then standard arguments using the fact that $\mathcal{K}'$ is Euclidean can be used to show $(M', s_0) \models \neg KK\psi'$. Thus, $K\psi' \in F_2$, and $(M, s_{\psi'}^{F_2}) \models \neg \psi'$ by the induction hypothesis. Again, it follows that $(M, s) \models \neg K\psi'$. ∎

The proof of Theorem 3.1 in the case that $C = \{e\}$ now follows easily. To check that $\varphi$ is satisfiable in $\mathcal{M}^e$, we simply guess a structure of the form described in Lemma 3.2 and verify that it does indeed satisfy $\varphi$. (Verifying that the structure guessed does indeed satisfy $\varphi$ is an instance of the model-checking problem, which is well known to be in polynomial time [Fagin, Halpern, Moses, and Vardi 1995]). Thus, the problem is in *NP*. The argument for the other cases with $e \in C$ follows by a straightforward modification of Lemma 3.2, as outlined just before the proof of the lemma. ∎

## References

Fagin, R., J. Y. Halpern, Y. Moses, and M. Y. Vardi (1995). *Reasoning about Knowledge*. Cambridge, Mass.: MIT Press. A slightly revised paperback version was published in 2003.

Halpern, J. Y. and Y. Moses (1992). A guide to completeness and complexity for modal logics of knowledge and belief. *Artificial Intelligence 54*, 319–379.

Ladner, R. E. (1977). The computational complexity of provability in systems of modal propositional logic. *SIAM Journal on Computing 6*(3), 467–480.

Vardi, M. Y. (1989). On the complexity of epistemic reasoning. In *Proc. 4th IEEE Symp. on Logic in Computer Science*, pp. 243–252.